\pdfoutput=1
\documentclass{article}[12pt]
\usepackage{epsf,amssymb,amsmath,latexsym}

\usepackage{tabularx}

\usepackage{color,graphicx}

\begin{document}

\thispagestyle{empty}

\begin{center}



{\large \bf
Testing Hawking particle creation by black holes \\ through correlation measurements
\footnote{March 31, 2010. Essay written for the GRF 2010 Awards for Essays on Gravitation.}
}
\bigskip
\bigskip
\bigskip





{ \sc R. Balbinot$^a$, I. Carusotto$^b$, A. Fabbri$^{c}$ and A. Recati$^{b}$ }
\footnote{balbinot@bo.infn.it, carusott@science.unitn.it, afabbri@ific.uv.es, recati@science.unitn.it}
\bigskip
\bigskip

{{\it a) } Dipartimento di Fisica dell'Universit\`a di Bologna and INFN sezione di Bologna,  Via Irnerio 46, 40126 Bologna, Italy\\
{\it b)} CNR-INFM BEC Center and Dipartimento di Fisica, Universit\`a di Trento, via Sommarive 14,
I-38050 Povo, Trento, Italy }\\
{\it c)} Departamento de Fisica Teorica and IFIC, Universidad de
Valencia-CSIC,  C. Dr. Moliner 50, 46100 Burjassot, Spain\\


\vspace*{1.5cm}

\large{\bf Abstract}
\end{center}
\noindent
Hawking's prediction of thermal radiation by black holes has been shown by Unruh to be expected also in condensed matter systems. We show here that in a black hole-like configuration realised in a BEC this particle creation does indeed take place and can be unambiguously identified via a characteristic pattern in the density-density correlations. This opens the concrete possibility of the experimental verification of this effect.

\bigskip
\bigskip


\newpage

\setcounter{page}{1}

Gravity is by far the weakest interaction in nature. Although it governs the large-scale structure of the
Universe,
its influence on microscopic physics is negligible.
Nevertheless, unlike the other interactions gravity is always attractive and this, under special circumstances, leads to the formation of objects whose gravitational field is so strong that not even light can escape. This was recognized already
in the 18th century by Mitchell and later by Laplace \cite{MiLa}, who predicted the existence of Newtonian dark stars, compact objects with an escape velocity bigger or equal to the velocity of light $c$. Their size is smaller or at most equal to the Schwarzschild radius $r_G=2GM/c^2$. A much more refined result comes from a general relativistic treatment of light propagation in a gravitational field, but the conclusion is essentially the same: were nature able to allow for objects collapsing through  $r=r_G$, we would have the formation of a region ($r<r_G$) invisible to an exterior observer: a black hole. The surface $r=r_G$ is called the event horizon, because it marks the outer boundary of the spacetime region which is not in causal contact with the outside world at all times.
The crucial feature of the event horizon, that we  will exploit later, is represented in Fig. 1. The future history of two nearby outgoing light rays, initially situated just outside and just inside the horizon, takes them completely apart: the exterior one (escaping) travels to infinity where it arrives highly redshifted, while the interior one (trapped) falls down towards the singularity.

Although according to General Relativity black holes are indeed black objects, this is no longer true when
Quantum Mechanics is taken into account.
Stephen Hawking in 1974
made the spectacular prediction that
black holes emit thermal radiation as a hot body \cite{Hawking:1974rv}.
He considered what happens to quantum fields
when a horizon is formed by the collapse of a star leading to a black hole.
The initial vacuum state $|0\rangle_{in}$ of the quantum field before the star starts to collapse
appears at late times, after horizon formation, as a two-mode squeezed vacuum state
\begin{equation}\label{inout}
|0 \rangle_{in} \propto \mbox{exp}(
\sum_\omega e^{
-\frac{\pi c \omega}{\kappa_B \kappa}}
a_\omega^{\dagger(esc)}a_\omega^{\dagger(trap)} )
|0\rangle_{esc} \times |0\rangle_{trap} \ ,
\end{equation}
where $\kappa$ ($=c^4/4GM$ for Schwarzschild black holes) is the horizon's surface gravity, $|0\rangle_{esc}$ is the vacuum of the escaping modes, $|0\rangle_{trap}$ the vacuum for the trapped modes
and $a_\omega^{\dagger(esc)}$, $a_\omega^{\dagger(trap)}$ the corresponding particle creation operators.
The tensorial structure reflects the fact that after the horizon formation the space-time is composed by two
causally disconnected regions, the black hole (BH) and the exterior (EXT) one.
Hawking 
considered measurements to be performed in the exterior region, where the inaccessible black hole degrees of freedom must be integrated out. Surprisingly he found that the black hole behaves as a black body emitting thermal radiation
at the temperature
\begin{equation}\label{tempe}
T_\mathrm{H}=\frac{\hbar \kappa}{2\pi k_B c}\ .
\end{equation}
This led to such a beautiful synthesis between gravity and thermodynamics (``The laws of black holes thermodynamics'')
that although not all aspects of the derivation are free of criticisms, as we shall discuss later, many people consider Hawking radiation as a milestone of the still to be discovered quantum theory of gravity.

Unfortunately Hawking's remarkable prediction has so far no experimental support, since for
a solar mass black hole $T_H\sim 10^{-7}\ K$, many orders of magnitude below
the cosmic microwave background temperature. For more massive black holes the emission temperature is even smaller. There is no hope to identify such a tiny signal in the sky. Furthermore there is no observational evidence of an excess of radiation on the $X$-ray cosmic background which could be associated to Hawking radiation from a primordial population of mini black holes with mass around $10^{15}\ g$.

As we shall see convincing evidence of the existence of Hawking radiation comes from a field which is surprisingly far away from gravity, namely condensed matter physics: a seminal work by Unruh \cite{Unruh:1980cg} pointed out the mathematical equivalence between the propagation of a massless scalar field in a curved spacetime and sound propagation in an inhomogeneous fluid, the so called gravitational analogy.  This led Unruh to infer the production of a thermal flux of phonons whenever in a fluid an acoustic horizon (i.e. a region separating subsonic from supersonic flow) is formed. The reasoning is the same as Hawking's, and rests on the fact that the nature of the effect is purely kinematical, i.e. it does not depend on the underlying dynamics but just on the identical features of wave propagation near a horizon whatever its origin, gravitational or sonic.

However both Hawking's and Unruh's derivation of the effect have a weak point that might invalidate the end result: they seem to rely on the propagation of very short wavelength modes near the horizon, subplanckian in the gravitational case and smaller than the intermolecular distance in the fluid case.
This is because of the exponential red (Doppler in the fluid case) shift suffered by the particles modes constituting Hawking radiation in their journey from the horizon region to infinity.
Only extremely short wavelength modes created near the horizon can reach infinity.

Extending the classical description of the space-time to Planckian scales, as implicit in Hawking's derivation, or the fluid description to scales smaller than the intermolecular distance seems highly hazardous and the very existence of Hawking radiation becomes questionable, since it might appear an artifact of the long wavelength approximation.

This is the so called ``transplanckian'' problem which seriously afflicts Hawking radiation \cite{Jacobson:1991gr}. There is no hope to address properly (not just through ad hoc toy models) this problem in the gravitational case since up to today no complete and consistent quantum theory of gravity is available for a description of space-time at the Planck scale. On the contrary, in the fluid case one often has a complete and well tested quantum mechanical microscopic description of the system at the molecular or atomic scale and the existence of Hawking radiation can then be investigated at a very fundamental level and the theoretical predictions tested in laboratory experiments.


Among the many systems proposed, Bose-Einstein condensates of ultracold atoms offer probably the most favourable experimental conditions. Indeed, the generally huge difference between the system temperature and the anticipated Hawking temperature of the emission can be dramatically reduced.
To further separate the Hawking-Unruh phonons from the background coming from thermal phonons, competing processes, and quantum noise one has to consider an aspect of eq. (\ref{inout}) that is usually neglected in the astrophysical context: Hawking particles are created in pairs, one in the escaping sector, which contributes to the thermal radiation at infinity, the other in the trapped sector, the ``partner'' which eventually gets swallowed by the singularity \cite{Brout:1995wp}.

The initial vacuum state contains local correlations between escaping and trapped modes which are converted, in the course of the time evolution, to nonlocal correlations between the $BH$ and $EXT$ regions, in the way represented in Fig. 1.
Such correlations, lost in the tracing operation associated to measurements in the exterior region, have a typical form that characterizes the Hawking effect.

Of course, the study of these correlations in gravitational black holes is of purely academic interest due to the impossibility of measuring them, since the BH region is inaccessible to the external world.
This is no longer true for condensed matter systems, where the acoustic origin of the horizon does not forbid correlation measurements between the BH and EXT regions.
In addition, measurements of quantum noise correlations are nowadays a well-established tool to experimentally address the microscopic physics of ultracold atomic gases~\cite{zw}.

According to the mean-field theory of dilute BECs \cite{PiSt}, the condensate can be described by a $c$-field satisfying the Gross-Pitaevski equation. Quantum fluctuations above the condensate are described by a quantum field satisfying the Bogoliubov-de Gennes equations. This field can be decomposed into density ($n_1$) and phase ($\theta_1$)
fluctuations.
On scales bigger than the healing length $\xi$
, a sort of Compton wavelength (which would correspond to the Planck scale in the gravitational case), one can use the hydrodynamical (fluid) approximation where for one-dimensional flow
\begin{equation}
\label{deph}
n_1=-\frac{\hbar}{g}(\partial_t\theta_1 + v\partial_x \theta_1)\ , \end{equation}
with $g$ the atom-atom nonlinear interaction constant and $v$ the condensate velocity, and $\theta_1$
satisfies an equation which is formally identical to a massless Klein-Gordon equation in a fictitious curved space-time metric (acoustic geometry)
\begin{equation}\label{metric}
ds^2=\frac{n}{mc(x)}\left[ -(c^2(x)-v^2)dt^2 -2vdtdx +dx^2 + dy^2 + dz^2 \right]\ ,
\end{equation}
where $y$, $z$ are the transverse dimensions and for simplicity the number density $n$ and velocity $v$ ($<0$) are kept constant ($m$ is the mass of the atoms) and the only non-trivial quantity is the speed of sound $c(x)$, tuned by varying $g$.

If the condensate is prepared in a way that the motion becomes supersonic in one region (say, $c(x)<|v|$ for $x<0$
and $c(x)>|v|$ for $x>0$) the formation of the sonic horizon at $x=0$ induces, according to Unruh's analysis
adapted to the actual system, the production of correlated pairs (particle-partner) of Bogoliubov phonons.
In the asymptotic $x>0$ region these phonons have a thermal distribution at the Hawking temperature (\ref{tempe}), where now the surface gravity is $\kappa=c\frac{dc}{dx}$ evaluated on the horizon $x=0$.
The signal of this pair-production can be found in the density-density correlation function,
that using (\ref{deph}) can be expressed simply by derivatives of the $\theta_1$ two-point function.
\noindent
If we consider points sufficiently far away so that the speed of sound is well approximated by its asymptotic values $c(x)\to c_l$, $c(x')\to c_r$  a straightforward calculation of the normalized density-density
in the $|0\rangle_{in}$ state at $t=t'$ gives \cite{Balbinot:2007de}
\begin{equation}\label{gtwo}
G^{(2)}(t;x,x')= \frac{\kappa^2 \xi_l\xi_r}{16\pi c_lc_r\sqrt{(n\xi_l)(n\xi_r)}} \frac{c_rc_l}{(v+c_l)(v+c_r)}\cosh^{-2}[\frac{\kappa}{2}(\frac{x}{v+c_l}-\frac{x'}{v+c_r})]\ . \ \ \ \ \ \ \ \ \ \ \ \  \end{equation}
These correlations characterize neatly the Hawking effect, namely the dependence on the surface gravity in the height and width and, much more important, the stationary peak at $\frac{x}{v+c_l}=\frac{x'}{v+c_r}$ identifying pairs of quanta created just inside and just outside the horizon at its formation and traveling in opposite directions. These conclusions are obtained considering the hydrodynamical (long wavelength) approximation of the BEC theory and exploiting the gravitational analogy.

However a key point is to know whether in a BEC Hawking radiation and its features are robust, i.e. if they survive when a detailed ``ab initio'' calculation with the full microscopic theory is performed overcoming the transplanckian problem. The definitively positive answer to this fundamental question has been given in \cite{Carusotto:2008ep} and can be clearly seen in Fig. 2, where the result of a numerical computation  based on the so called truncated Wigner method are presented.
The spectacular appearance after the horizon formation of the two tongues neatly peaked exactly along the line
$\frac{x}{v+c_l}=\frac{x'}{v+c_r}$ as predicted by the gravitational analogy signals without doubts the actual presence of Hawking radiation in BECs. It has been also shown that the effect is still present and clearly visible even in the presence of a thermal background, always present in BECs.

This is the first firm evidence that Hawking radiation is a true physical effect and we expect it to be quite general.
Quantitatively the deviations of the exact result from the simple expression (\ref{gtwo}) obtained via the gravitational analogy disappear when the scale of variation of the sound velocity profile across the horizon is sufficiently larger than the healing length.
Inserting realistic figures for existing experiments~\cite{atomlaser}, correlations of the order of $10^{-3}$ can be anticipated. This value is small but not far away from the sensitivity of actual experiments. Furthermore, proposals to amplify the signal by a factor 10 by following the formation of the acoustic black hole with a period of free expansion \cite{cornell09} may be important for the experimental search. These facts, together with the growing progress in the technology, indicate that density correlation measurements are the most promising way for a direct experimental verification of the Hawking-Unruh effect in the near future.


This remarkable result makes us confident that even for the usual gravitational black holes Hawking radiation indeed exists, and that the semiclassical result (\ref{inout}) is valid up to scales of the order of the Planck length. This is relevant if mini black holes will be really produced at LHC to properly identify their signal.
Moreover, we may speculate that provided black holes evaporate completely according to unitary rules then the partners must eventually come out of the horizon. This would allow to find peculiar correlations performing measurements at early and late times.



\newpage

\begin{figure}[h]
\centering \includegraphics[angle=0,width=\textwidth, bb= 25 25 400 400] {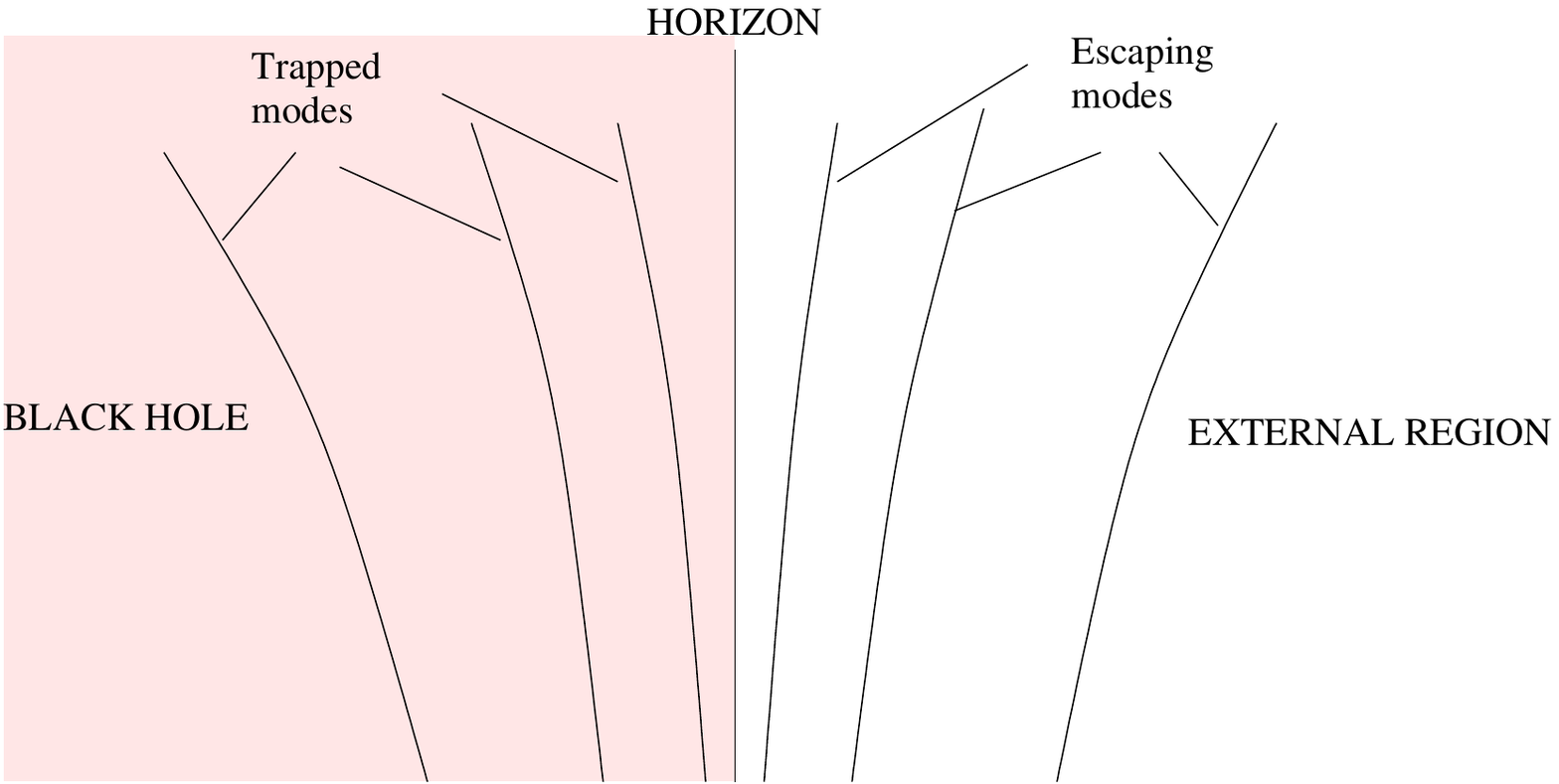}
\caption{\label{uno} Escaping and trapped outgoing null rays in a black hole spacetime.}
\end{figure}

\begin{figure}[h]
\centering \includegraphics[width=\textwidth, bb= 20 20 500 500]{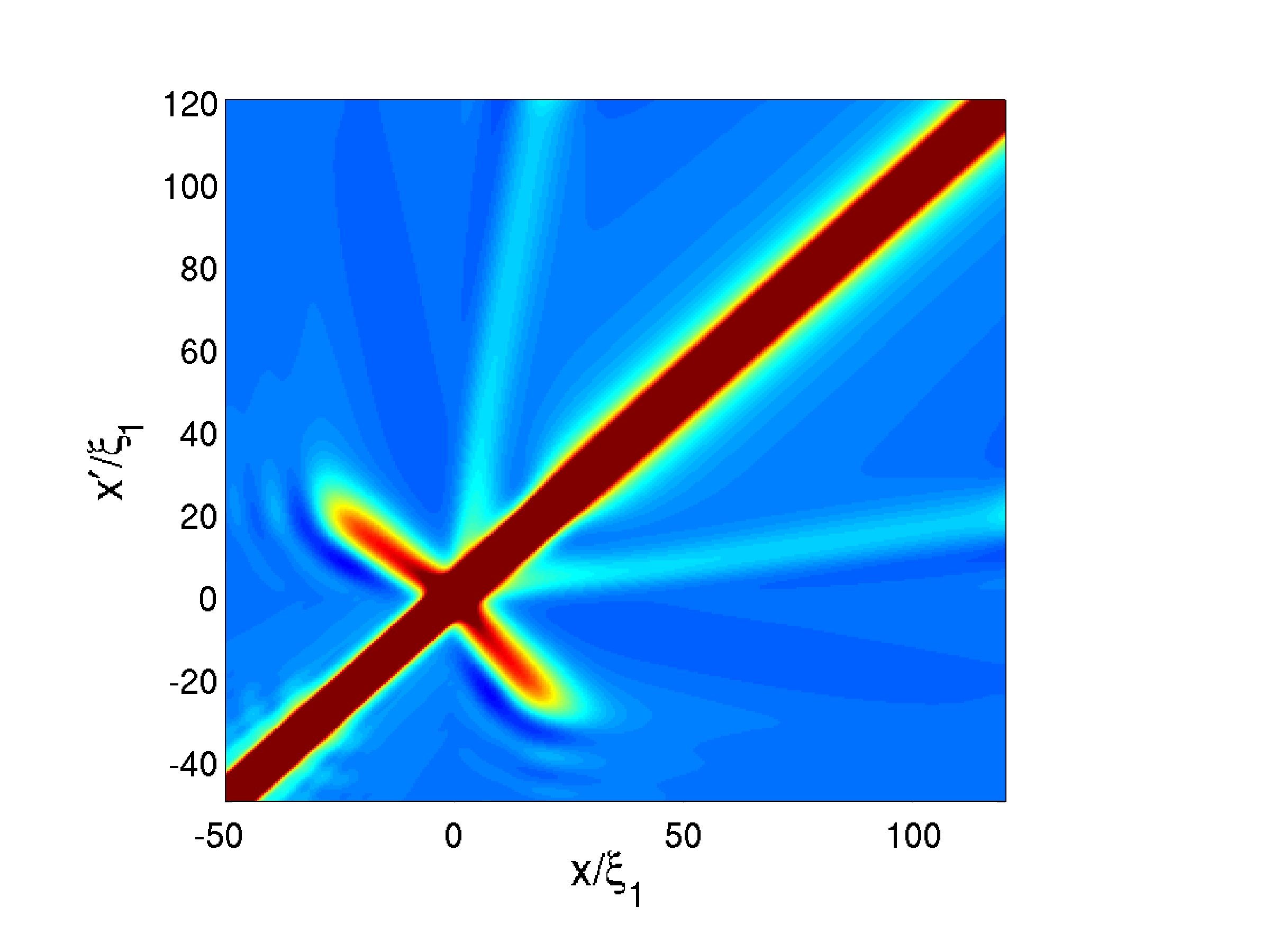}
\caption{\label{uno} Example of density plot of the normalized correlation function of density fluctuations
$G^{(2)}(x,x')=(n\xi) <n_1(x) n_1(x')> / n^2$. The color scale goes from red ($-8\times
10^{-3}$) to light blue ($2\times 10^{-3}$). The main Hawking signal is the red tongue
oriented along the NW-SE straight line.}
\end{figure}

\end{document}